# Leveraging Adaptive Model Predictive Controller for Active Cell Balancing in Li-ion Battery

Seyed Mahmoud Salamati [a], Seyed Ali Salamati [b], Mohsen Mahoor [c], and Farzad Rajaei Salmasi [d]

North Carolina State University [a], University of Tehran [b,d], University of Denver [c]

ssalama@ncsu.edu, salamati@ut.ac.ir, Mohsen.Mahoor@du.edu, Farzad_rs@ieee.org

*Abstract*— **Automotive industry is moving toward fully electric and hybrid electric vehicles. Accordingly, energy storage unit is one of the most important blocks in these electric drives. Battery stacks which contain a number of cells are being used for supplying the vehicles' energy. Charge equalization for series connected battery strings has a significant effect on battery life. In this paper, an adaptive model predictive controller (AMPC) is proposed to manage the cell equalizing process. The series connected cells' voltages and currents are collected, then leveraging Recursive Least Square (RLS) method, the future voltage samples for all of the cells are predicted. MPC controller specifies a sequence which results in the optimum balancing performance of the proposed circuit. Simulation results prove that using the suggested algorithm, the voltage set of the series cells has moved more uniformly.**

I. INTRODUCTION

Electric Vehicles (EVs) and Hybrid Electric Vehicles (HEVs) are considered to be the future main transportation choices. Batteries are the most prominent energy sources for supplying the electric motor power demand in these cars [1]. Moreover, batteries are becoming an integral part for distributed generation schemes [2-3]. Different types of the batteries, such as Lead-acid, Ni-cd and Li-ion cells are being used in today's electric transportation vehicles. Because of limited voltage of an individual cell, commercial cars that use batteries as one of their energy supplies, have stacks which are comprised of several series and parallel cells. These battery stacks usually are designed to have voltages more than 250 Volts [4]. Considering high voltage per cell, power and energy density besides low internal resistance and self-discharge rate, Li-ion cells will be the first choice for the next generation of EVs. Due to special chemistry of Li-ion cells, they need to have a well-designed battery management unit. The most important task of Battery Management System (BMS) unit is to prevent individual cells' voltage to go under or beyond the lower or upper bands [5]. The charging controller does not take every cell into consideration and assumes the whole stack as a single battery with a specified voltage and capacity.

The most common strategy for charging Li-ion batteries is the constant current-constant voltage (CC-CV) method [6]. During both charge and discharge, voltage of every individual cell is sensed and if it is near to the forbidden area (above top voltage limit or below lower voltage limit of the Li-ion cell), charging or discharging will be stopped. Assuming that, initially, all of the series cells are at the same State of the Charge (SOC), a voltage (and SOC) drift will appear among them after hundreds of charge and discharge cycles, due to the construction differences, and non-uniform temperature gradient [7-9]. During the charge and discharge cycles, SOC and hence voltage of series connected cells will vary such that some of the cells hit one of the two upper or lower limits sooner than others. As a result, the whole capacity of the stack will be specified by the cells which are most near to the upper and lower bands [10]. The process which aims at bringing voltage and SOC equality among cells connected in series in a stack, is called cell equalization.

Cell equalizing methods are divided into two main categories: passive and active cell balancing [11]. In passive balancing solutions, the additional energy of the overcharged cells will be dissipated in resistive branches; so, they are not efficient, especially for usage in EVs because of Low efficiency. In active voltage equalizing methods, additional charge of cell(s) with higher voltage, will be transferred to the cell(s) with lower voltage. Several schemes are developed in order to maximize the balancing speed and also useful capacity of the stack. A comprehensive review of these methods can be found in [12] and [13]. Active cell balancing methods are divided into central balancing and adjacent-cell based on equalizing schemes.

The main contribution of this paper is introducing a model predictive approach which enhances quality of cell equalization process by making a uniformity between voltage gaps among stack cells during whole of the process. In contrast with conventional balancing algorithms which try to remove voltage gaps one after another, in this approach, voltage differences among cells are decreased uniformly. Explanation is that, electric circuit model of Li-ion batteries consists of one or more resistive-capacitive branches. When the cell with highest SOC is chosen for discharging in many sequential cycles, those (modelling) capacitors will be (negatively) charged; which results in decreasing terminal voltage such that internal SOC is not reflected and hence makes cell choosing wrong. The algorithm that is suggested here, is to optimally discharge cells with additional charge such that voltage gap (and hence SOC difference) of group of cells vary uniformly. Therefore, because of remaining a detectable voltage gap between stack cells, the cell which should be chosen in order to transfer its energy to the stack is selected more reliably.

A key point about the suggested algorithm is its dependence on Li-ion cell model. Adaptive schemes can be used for considering parameter variations of the model [14-15]. Here, recursive least square is used in order to deal with SOC dependency of parameters of Li-ion battery. The identified model of individual cells is then used in order to predict future cell voltage.

The remaining of the paper is organized as follows. Section II describes the online Li-ion cells' identification method. Structure of the balancing circuit and its operational stages are presented in Section III. Simulation results are provided in Section IV to verify the effectiveness of the proposed method. Finally, Sections V concludes the paper.

## II. ONLINE IDENTIFICATION OF CELLS' MODEL

In the proposed algorithm for cell balancing, the control action is based on the model of the cells in the stack and also the model of the convertor. The model of flyback convertor is not complex and will be discussed in the following sections. Li-ion battery can be modeled by chemical, mathematical or electrical models [16]. Chemical models are based on partial differential equations and are the best choice during design and production of batteries [17]. Mathematical models such as neural network based models include so many parameters and weights and are not proper for control goals [18]. The electrical models, besides on relative simplicity, are accurate enough [19]. [20] suggests a combined electrical model for Li-ion batteries. This model incorporates two resistive-capacitor branches, an internal resistor, open-circuit voltage which is shown by a controlled voltage source and model of self-discharging of the battery. The following equations describe mentioned electric circuit model in the state space form.

$$\begin{aligned} \dot{SOC} &= -\frac{1}{R_{sd}C_b}SOC - \frac{1}{C_b}I_B \\ \dot{V_1} &= -\frac{1}{R_1 C_1}V_1 + \frac{I_B}{C_1} \\ \dot{V_2} &= -\frac{1}{R_2 C_2}V_1 + \frac{I_B}{C_2} \\ \dot{V_B} &= V_{oc}(SOC) - V_1 - V_2 - R_0 I_B \end{aligned} \quad (1)$$

The open-circuit voltage of Li-ion battery as a function of state of the charge is given by:

$$\begin{aligned} V_{oc}(SOC) &= \alpha_0 + \alpha_1 SOC + \alpha_2 SOC^2 + \alpha_3 SOC^3 \\ &+ \alpha_4 e^{-\beta SOC} \end{aligned} \quad (2)$$

For controlling the cell equalizing process, we do not need to consider the self-discharging because it happens slowly. The remained battery model's parameters are changing both in short-term and long-term [21]. Therefore, a model which is established at a certain time cannot show an acceptable performance forever. Recursive least square errors (RLS), can be an efficient identification method if the proper regressor vector is chosen.

Considering the equations (1) and (2), it can be said that the terminal voltage of a Li-ion battery is a function of SOC and the terminal current. Moreover, existence of $\alpha_0$ in equation (2) shows that a constant value is a part of non-linearity of the battery model. So the voltage of the battery can be expressed as follows:

$$V_i(k) = f([I_i(k), \frac{\int I_i(\tau)\,d\tau}{C_b}, 1]) \quad (3)$$

If we take $X_i(k) = [C_2 I_i(k), C_1 \frac{\int I_i(\tau)d\tau}{C_b}, C_0]$ as the regressor vector for the ith cell in the stack and $\theta_i(k)$ as the desire weight vector, the identification steps are shown in Fig. 1.

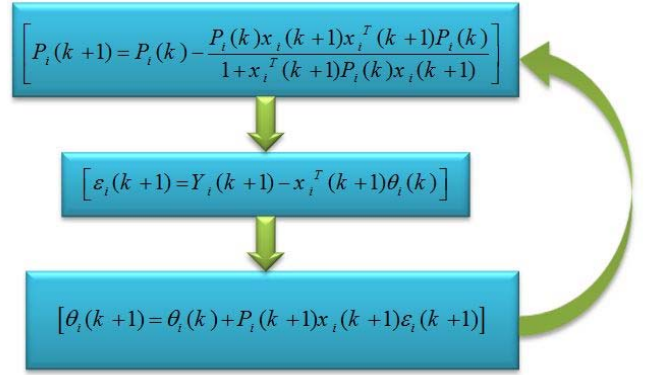

Fig. 1: Identification steps in RLS Algorithm

## III. PROPOSED TOPOLOGY AND ALGORITHM

The suggested cell balancing method can be categorized among the central equalizers because of using a multi winding flyback convertor as its key component (Fig. 2). This isolated DC-DC convertor, shares the energy between cells of the stack according to the sequence that is specified by the central controller. Taking the individual cells' and converters' model into consideration is the key feature of the proposed cell equalizing method that distinguishes it from the former algorithms. In this section, the basic principles of the proposed method are described with considering a stack of four series connected cells.

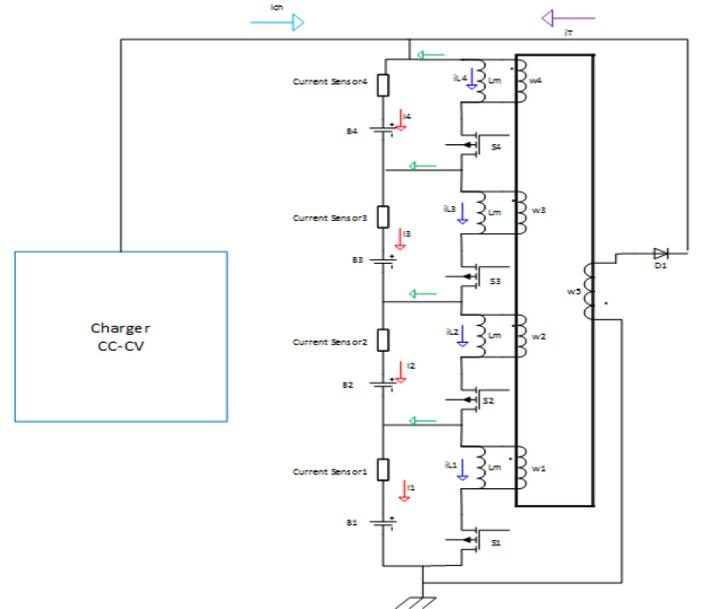

Fig. 2. The proposed circuit for cell balancing control

At the beginning of each balancing cycle, voltages of every cell are measured and read by the central controller. The cells in the stack will be sorted according to their voltage, from the highest to the lowest. If difference between the first and the fourth cell is more than 0.02 Volts, the MCU decides to enter the stack into the equalizing process. The controller at any cycle chooses the cell with the highest voltage to be discharged through its corresponding switch. The switch will remain at ON state until the current reaches to a predefined threshold. The time that is needed for the current of the highest cell's winding to reach to the peak value can be calculated using the following formula:

$$T_{on} = L \frac{I_{max}}{V_{cell}(t)} \quad (4)$$

This time interval can be divided into two equal windows which we call them stage (I) and (II). At both of these stages, the highest cell's corresponding switch is ON. Any of the second and third cells can be chosen by the controller to be discharged at any of the stages (I) and (II).

As it was stated, at the first of any cycle, for every cell, an updated model is produced. Using the produced model of the cells, the controller will be able to predict the behavior of the equalizing process under any combination of the switches' state. In fact, the controller decides that which of the second and the third cell will conduct in the first and the second stages (Table. I). The decision will be made according to the generated standard deviation (*std*) of four cells at the end of the next balancing cycle. The set of states in the first two stages, which result in the least *std*, will be chosen by the central controller (Fig. 2).

$$i_{L1}(t) = \frac{1}{L_m} \int_{t_0}^{t} V_{B1}(\tau)\, d\tau$$

$$i_{L2}(t) = C_1(1) * \frac{1}{L_m} \int_{t_0}^{t} V_{B2}(\tau)\, d\tau$$

$$i_{L3}(t) = C_2(1) * \frac{1}{L_m} \int_{t_0}^{t} V_{B3}(\tau)\, d\tau$$

$$i_{L1}(t) = 0 \quad (5)$$

In the above equations, $i_{Li}(t)$ and $V_{Bi}(t)$ shows the current of the magnetizing branch of $i^{th}$ winding in the flyback convertor model and $i^{th}$ cell's voltage. $C_1(1)$ and $C_2(1)$ are 1 if the controller has decided to turn the 2nd or the 3rd highest voltage cell's corresponding switch to be on in the first stage, or 0 if they are decided to be in the off state. This stage finishes when the controller's second sample time is started. Fig. 3 shows the different circuit diagrams that balancing process would experience under different control decisions.

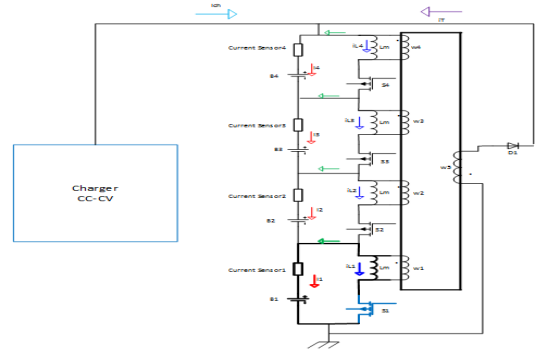

(a)

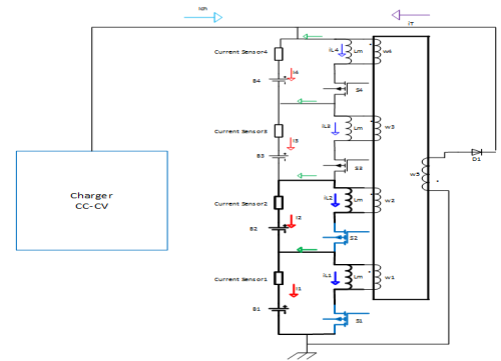

(b)

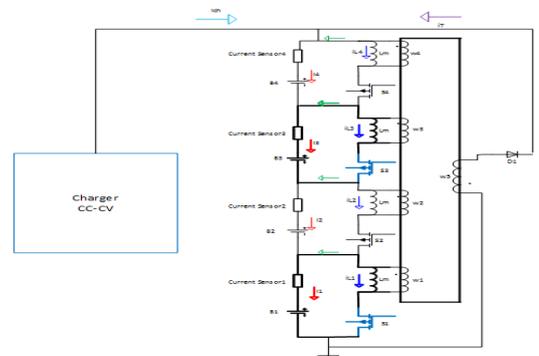

(c)

TABLE I
THE SEARCH SPACE FOR THE MODEL PREDICTIVE CONTROLLER

| Switch | $T_1$ | $T_2$ | Switch | $T_1$ | $T_2$ | Std |
|---|---|---|---|---|---|---|
| $S_1$ | OFF | OFF | $S_2$ | OFF | OFF | Std1 |
| $S_1$ | OFF | OFF | $S_2$ | OFF | ON | Std1 |
| $S_1$ | OFF | OFF | $S_2$ | ON | OFF | Std1 |
| $S_1$ | OFF | OFF | $S_2$ | ON | ON | Std1 |
| $S_1$ | OFF | ON | $S_2$ | OFF | OFF | Std1 |
| $S_1$ | OFF | ON | $S_2$ | OFF | ON | Std1 |
| $S_1$ | OFF | ON | $S_2$ | ON | OFF | Std1 |
| $S_1$ | OFF | ON | $S_2$ | ON | ON | Std1 |
| $S_1$ | ON | OFF | $S_2$ | OFF | OFF | Std1 |
| $S_1$ | ON | OFF | $S_2$ | OFF | ON | Std1 |
| $S_1$ | ON | OFF | $S_2$ | ON | OFF | Std1 |
| $S_1$ | ON | OFF | $S_2$ | ON | ON | Std1 |
| $S_1$ | ON | ON | $S_2$ | OFF | OFF | Std1 |
| $S_1$ | ON | ON | $S_2$ | OFF | ON | Std1 |
| $S_1$ | ON | ON | $S_2$ | ON | OFF | Std1 |
| $S_1$ | ON | ON | $S_2$ | ON | ON | Std1 |

*A. First stage*

Before starting of this stage, no one of the windings contained current. Therefore, during the first stage, no current will flow through the diode in the secondary winding; so, the balancing current of the stack will remain at zero. The general form of the equations which demonstrate variations of the different inductors' currents can be expressed as the following relationships.

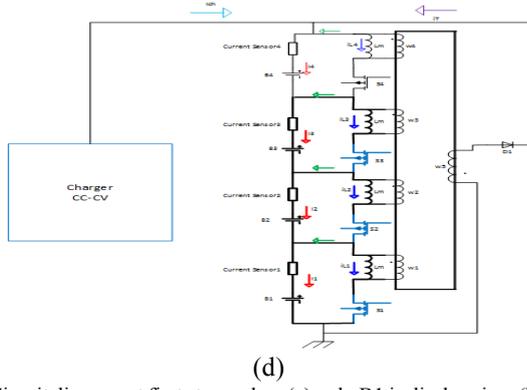

(d)

Fig. 3. Circuit diagram at first stage when (a) only B1 is discharging. (b) B1 and B2 are discharging. (c) B1 and B2 are discharging. (d) B1, B2 and B3 are discharging

*B. Second Stage*

In the second stage, the current in the highest cell's winding continues to increase. If state of second and the third cells' switches during the last stage were off, even if both switches keep their last state, with consideration of new initial conditions, the circuit operates in the same manner that set of equation (3) express. The secondary winding will conduct in a situation at which, at least one of two corresponding switches of the 2nd and the third change its state from on to off.

$$
\begin{aligned}
i_{L1}(t) &= i_{L1}(t_1) + \frac{1}{L_m}\int_{t_1}^{t} V_{B1}(\tau)\,d\tau \\
i_{L2}(t) &= i_{L2}(t_1) + \frac{C_1(2)}{L_m}\int_{t_1}^{t} V_{B2}(\tau)\,d\tau - \frac{(1-C_1(2))}{zz} \times \\
&\quad \int_{t_1}^{t}\Big[\frac{N_1}{N_2}(V_{B1}(\tau)+V_{B2}(\tau)+V_{B3}(\tau)+V_{B4}(\tau)) \\
&\quad -V_{B1}(\tau)-V_{B3}(\tau)\times C_2(2)\Big]d\tau \\
i_{L3}(t) &= i_{L3}(t_1) + \frac{C_3(2)}{L_m}\int_{t_1}^{t} V_{B3}(\tau)\,d\tau - \frac{(1-C_2(2))}{zz} \times \\
&\quad \int_{t_1}^{t}\Big[\frac{N_1}{N_2}(V_{B1}(\tau)+V_{B2}(\tau)+V_{B3}(\tau)+V_{B4}(\tau)) \\
&\quad -V_{B1}(\tau)-V_{B3}(\tau)\times C_1(2)\Big]d\tau \\
i_{L4}(t) &= 0 \qquad (6)
\end{aligned}
$$

In the above equations, $C_1(2)$ and $C_2(2)$ can be defined similar to $C_1(1)$ and $C_2(1)$ but in the second stage. $N_1$ and $N_2$ demonstrate the number of turns in primaries and secondary of the flyback convertor. The *zz* parameter shows the number of the switches which have changed their state from off to on. Fig. 4 shows different possibilities that can influence the currents of the balancing circuit. The current which follows in the secondary side of the transformer is given by:

$$i_T(t) = \frac{[(1-C_1(2))\times i_{L2}(t) + (1-C_2(t))\times i_{L3}(t)]}{zz+2} \times \frac{N_1}{N_2} \quad (7)$$

Also, the balancing current which passes through different cells will vary in a manner that bellow equations can describe it.

$$
\begin{aligned}
i_{bal1}(t) &= -i_{L1}(t) + i_T(t) \\
&\quad -\frac{[(1-C_1(2))\times i_{L2}(t) + (1-C_2(t))\times i_{L3}(t)]}{zz+2} \\
i_{bal2}(t) &= i_T(t) - i_{L2}(t)\times C_1(2) \\
&\quad -C_1(2)\times \frac{[1-C_2(2)]\times i_{L3}(t)}{zz+2}
\end{aligned}
$$

$$
\begin{aligned}
i_{bal3}(t) &= i_T(t) - i_{L3}(t)\times C_2(2) \\
&\quad -C_2(2)\times \frac{[1-C_1(2)]\times i_{L2}(t)}{zz+2} \\
i_{bal4}(t) &= i_T(t) \qquad (8)
\end{aligned}
$$

where, *nn* shows the number of primary windings which still contain current. During this step, all of the balancing currents will be same as each other and equal to the secondary current ($i_T$) which is given by:

$$i_T(t) = [i_{L1}(t)+i_{L2}(t)+i_{L3}(t)]\times \frac{N_1}{N_2} \quad (9)$$

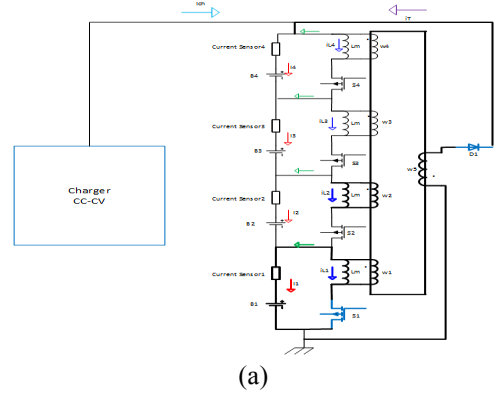

(a)

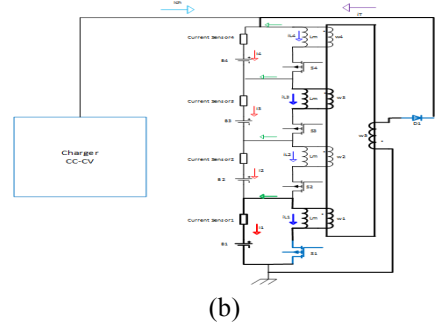

(b)

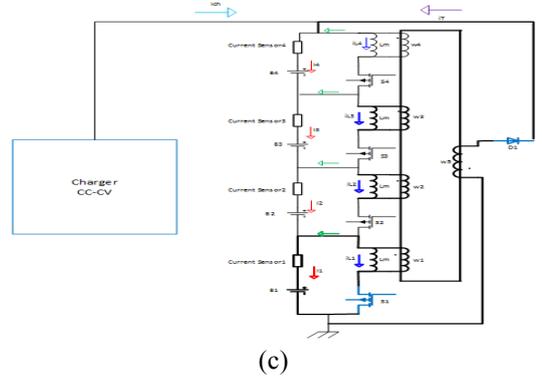

(c)

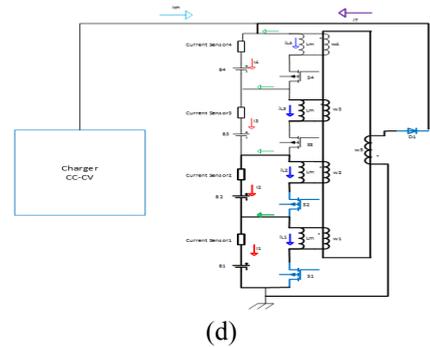

(d)

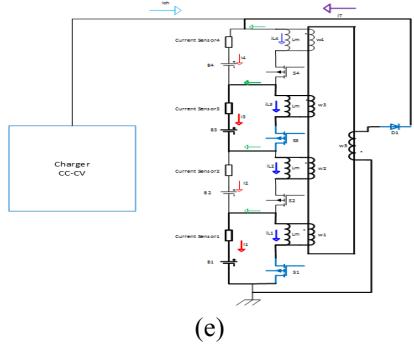

(e)

Fig. 4. Circuit diagram at the second stage when (a) B1 is discharging and B2 has gone from on to off state (b) B1 is discharging and B3 has gone from on to off state (c) B1 is discharging and B2 and B3 have gone from on to off state (d) B1 and B2 are discharging and B3 has gone from on to off state (e) B1 and B3 are discharging and B2 has gone from on to off state

*C. Third Stage*

By finishing the second stage, all the switches will remain in off state until the current of the secondary reaches to zero. The circuit diagram of this stage is shown in Fig. 5. So, the magnetizing current of different cells is given by:

$$
\begin{aligned}
i_{L1}(t) &= i_{L1}(t_2) - \frac{\int_{t_2}^{t} \left[\frac{N_1}{N_2}(V_{B1}(\tau)+V_{B2}(\tau)+V_{B3}(\tau)+V_{B4}(\tau))\right]}{nn} \\
i_{L2}(t) &= i_{L2}(t_2) - \frac{\int_{t_2}^{t} \left[\frac{N_1}{N_2}(V_{B1}(\tau)+V_{B2}(\tau)+V_{B3}(\tau)+V_{B4}(\tau))\right]}{nn} \\
i_{L3}(t) &= i_{L3}(t_2) - \frac{\int_{t_2}^{t} \left[\frac{N_1}{N_2}(V_{B1}(\tau)+V_{B2}(\tau)+V_{B3}(\tau)+V_{B4}(\tau))\right]}{nn} \\
i_{L4}(t) &= i_{L4}(t_2) - \frac{\int_{t_2}^{t} \left[\frac{N_1}{N_2}(V_{B1}(\tau)+V_{B2}(\tau)+V_{B3}(\tau)+V_{B4}(\tau))\right]}{nn}
\end{aligned}
$$
(10)

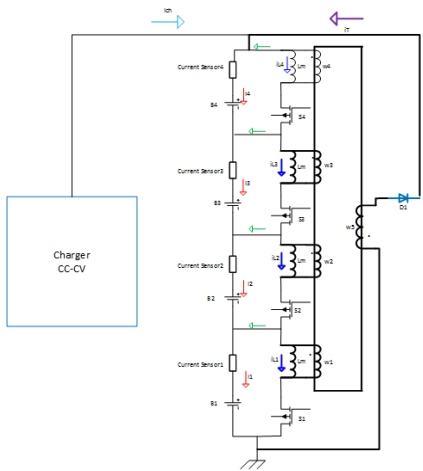

Fig. 5. Circuit diagram at the third stage

## IV. SIMULATION RESULTS

In order to demonstrate performance of the proposed cell balancing control method, the behavior of the balancing circuit is simulated in MATLAB environment. It is assumed that stack contains four cells with the same in capacity and other characteristics except the initial state of the charge. The cells' initial SOC are demonstrated in the table II.

TABLE II
INITIAL SOCs OF THE CELLS

| | |
|---|---|
| $SOC_1(0)$ | 60% |
| $SOC_2(0)$ | 50% |
| $SOC_3(0)$ | 45% |
| $SOC_4(0)$ | 40% |

The relative number of turns of the primary windings to the secondary winding is set 1 to 4. Also, all of the cells' electrical models' parameters are assumed to be the same as the parameters that are achieved for a 800 mAh Li-ion cell in [7]. The stack is assumed to be under the CC-CV charging process. The variations of the cell's SOCs, using the described cell balancing adaptive MPC controller are depicted in Fig. 6.

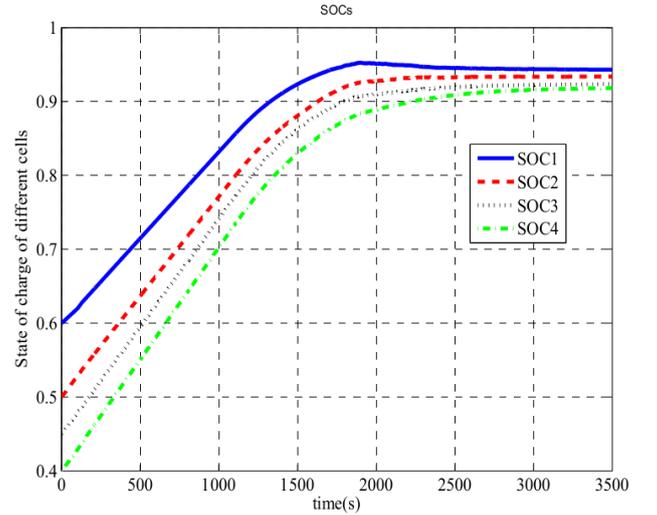

Fig. 6. The variation of SOCs of different cells using adaptive MPC cell balancing method

As it is obvious, SOCs of the different cells have moved uniformly during the balancing time. Figure (7) depicts the balancing currents of the different cells and figure (8) shows the voltages of the highest and lowest cells during the voltage equalizing process.

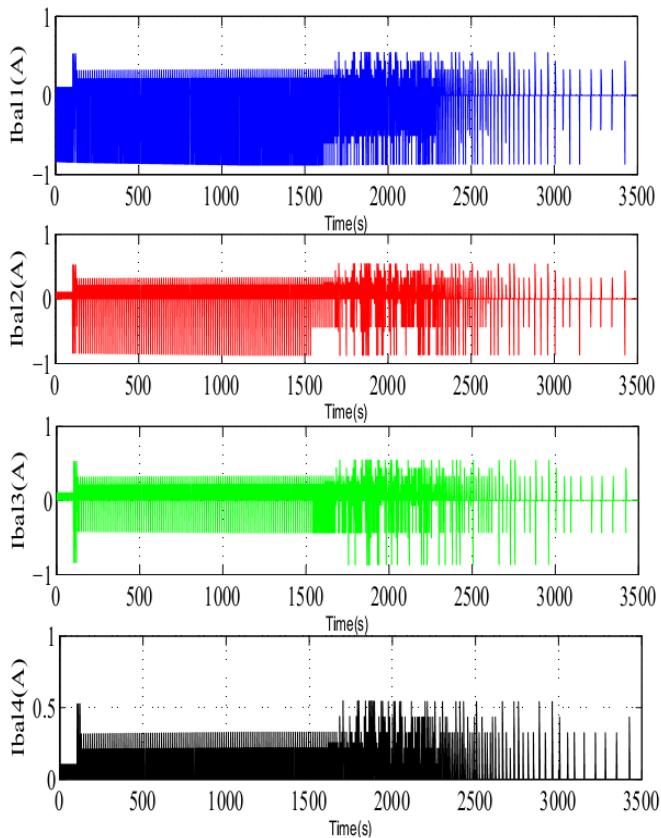

Fig. 7. Variations of different cells' balancing current

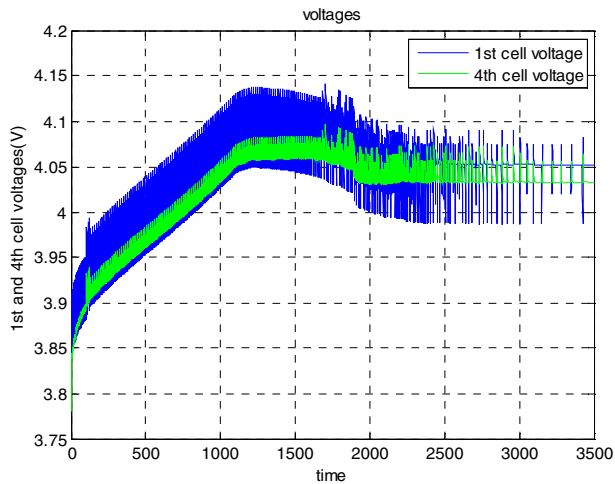

Fig. 8. Variations of voltage of the first and the fourth cells (initially highest and lowest ones)

## V. CONCLUSION

In this study, leveraging adaptive model-based predictive controller, an efficient approach for regulating the cell balancing in Li-ion battery was proposed. In this respect, the currents and the voltages of the cells were measured, then the future voltages of the cells were predicted using RLS identification method. Using the predicted voltages, the optimum sequence was selected for different switches of the designed circuit. Simulation results proved that applying the selected sequence to the switches will make the balancing process more efficient.